\documentstyle[11pt,paspconf,epsf]{article}

\def\degree{\ifmmode {^\circ}\else {$^\circ$}\fi}
\def\rstar{\ifmmode {R_{\star}}\else $R_{\star}$\fi}
\def\rsun{\ifmmode {\rm R_{\odot}}\else $\rm R_{\odot}$\fi}
\def\rsunsq{\ifmmode {\rm R_{\odot}^2}\else $\rm R_{\odot}^2$\fi}
\def\mstar{\ifmmode {M_{\star}}\else $M_{\star}$\fi}
\def\lstar{\ifmmode {L_{\star}}\else $L_{\star}$\fi}
\def\tstar{\ifmmode {T_{\star}}\else $T_{\star}$\fi}
\def\mearth{\ifmmode {\rm M_{\oplus}}\else $\rm M_{\oplus}$\fi}
\def\msun{\ifmmode {\rm M_{\odot}}\else $\rm M_{\odot}$\fi}
\def\msunyr{\ifmmode {\rm M_{\odot}\,yr^{-1}}\else $\rm M_{\odot}\,yr^{-1}$\fi}\def\mdot{\ifmmode {\dot{M}}\else $\dot{M}$\fi}
\def\mssunyr{\ifmmode {\rm M_{\odot}^2\,yr^{-1}}\else $\rm M_{\odot}^2\,yr^{-1}$\fi}\def\mdot{\ifmmode {\dot{M}}\else $\dot{M}$\fi}
\def\lsun{\ifmmode {\rm L_{\odot}}\else $\rm L_{\odot}$\fi}
\def\lbol{\ifmmode {L_{bol}}\else $L_{bol}$\fi}
\def\teff{\ifmmode {T_{eff}}\else $T_{eff}$\fi}
\def\ne{\ifmmode {n_{e}}\else $n_{e}$\fi}
\def\te{\ifmmode {T_{e}}\else $T_{e}$\fi}
\def\cm3{\ifmmode {\rm cm^{-3}}\else $\rm cm^{-3}$\fi}
\def\emm{\ifmmode {n_e^2 V}\else $n_e^2 V$\fi}
\def\gcm3{\ifmmode {\rm g~cm^{-3}}\else $\rm g~cm^{-3}$\fi}
\def\ergg{\ifmmode {\rm erg~g^{-1}}\else $\rm erg~g^{-1}$\fi}
\def\ergs{\ifmmode {\rm erg~s^{-1}}\else $\rm erg~s^{-1}$\fi}
\def\ecs{\ifmmode {\rm erg~cm^{-2}~s^{-1}}\else $\rm erg~cm^{-2}~s^{-1}$\fi}
\def\mum{\ifmmode {\rm \mu {\rm m}}\else $\rm \mu {\rm m}$\fi}
\def\nh3{\ifmmode {\rm NH_3}\else $\rm NH_3$\fi}
\def\arcsec{\ifmmode ^{\prime \prime}\else $^{\prime \prime}$\fi}

\def\inch{\ifmmode ^{\prime \prime}\else $^{\prime \prime}$\fi}
\def\arcmin{\ifmmode ^{\prime}\else $^{\prime}$\fi}

\def\lfl{\ifmmode {\lambda F_{\lambda}}\else $\lambda F_{\lambda}$\fi}
\def\lFl{\ifmmode {\lambda F_{\lambda}}\else $\lambda F_{\lambda}$\fi}
\def\footspace{\baselineskip=10pt}
\def\singlespace{\baselineskip=12pt}

\newbox\grsign \setbox\grsign=\hbox{$>$} \newdimen\grdimen \grdimen=\ht\grsign
\newbox\simlessbox \newbox\simgreatbox
\setbox\simgreatbox=\hbox{\raise.5ex\hbox{$>$}\llap
     {\lower.5ex\hbox{$\sim$}}}\ht1=\grdimen\dp1=0pt
\setbox\simlessbox=\hbox{\raise.5ex\hbox{$<$}\llap
     {\lower.5ex\hbox{$\sim$}}}\ht2=\grdimen\dp2=0pt
\def\simgreat{\mathrel{\copy\simgreatbox}}
\def\simless{\mathrel{\copy\simlessbox}}
\def\gtrsim{\simgreat}

\def\etal{{\it et al. }}

\begin{document}

\title{Dynamical Evolution of Protoplanetary Disks}


\author{Scott J. Kenyon}
\affil{Smithsonian Astrophysical Observatory, 60 Garden Street, Cambridge, MA 02138 USA}

\begin{abstract}
This paper reviews the theory of protostellar debris disks.  After 
a brief introduction to accretion disk physics, I describe coagulation 
models of planet formation in the outer regions of planetesimal disks.  
Coagulation models for the Kuiper Belt produce Pluto-sized objects on 
timescales of 10--40 Myr.  These models yield size distributions which 
agree with observations of Kuiper Belt objects with red magnitudes,
R $\approx$ 20--27.  Velocity stirring models for other debris disk
systems demonstrate that 500 km or larger objects can stir the velocities
of small objects up to the shattering limit needed to begin a collisional
cascade.  

\end{abstract}

\section{Introduction}

In the past two decades, several remarkable discoveries have radically 
altered our understanding of planet formation.  In 1983, data from the 
{\it Infrared Astronomical Satellite} revealed substantial mid-infrared 
emission from three A-type main-sequence stars, $\alpha$ Lyr, 
$\alpha$ PsA, and $\beta$ Pic (Aumann \etal 1983).  In addition to the 
flux expected from an A-type stellar photosphere, the observations
implied a cold, extended source of radiation.  These data led to 
a beautiful ground-based image of an edge-on dusty disk surrounding 
$\beta$ Pic (Smith \& Terrile 1984).  
This image provided dramatic evidence that the building blocks for 
solar systems exist in disk-like structures around other stars.  More 
recent images with the {\it Hubble Space Telescope} and large ground-based 
telescopes have revealed disks with similar sizes, 100--1000 AU, around 
many pre-main sequence and main sequence stars 
(e.g., Jayawardhana \etal 1998; Koerner \etal 1998;
Greaves \etal 1998; Augereau \etal 1999; Schneider \etal 1999;
see also the reviews by Beckwith and Koerner in this volume).  
As in $\alpha$ Lyr and $\beta$ Pic, all of these systems have excess 
near-infrared or mid-infrared emission compared to a normal stellar 
photosphere.  More than 50\% of young stars ($\sim$ 1 Myr) in nearby 
star-forming regions also have excess infrared emission, suggesting that
most stars are born with disks with sizes and masses at least as large 
as our solar system (e.g., Lada 1999; Jayawardhana \etal 1999).  

Ground-based radial velocity surveys have recently detected $\sim$ 
50 Jupiter-mass and Saturn-mass planets in close orbits around 
$\sim$ 1\% of nearby solar-type stars (e.g., Latham \etal 1989; 
Marcy \& Butler 1996, 1999; Cochran \etal 1997; Noyes \etal 1997;
DelFosse \etal 1998).  Several systems have multiple planets.  
Although some of these planet candidates may have masses close to 
the hydrogen-burning 
limit for main sequence stars,  these discoveries demonstrate that 
gas giant planets are relatively common in the solar neighborhood.
However, no planet has yet been unambiguously detected in a debris 
disk system.  Rings, warps, and other spatially extended structures
in some images suggest the tidal influence of planets embedded
in several debris disks (e.g., Kalas \& Jewitt 1995), but direct 
detection of planetary motion has proved difficult.  

The final interesting discovery links extra-solar disks and planets
with the formation and evolution of our solar system.  Recent surveys
have detected a substantial population of Kuiper Belt objects (KBOs),
small icy bodies outside the orbit of Neptune (Luu \& Jewitt 1988, 
1998; Luu \etal 1997; Jewitt \& Luu 1993). For an adopted albedo of 
4\%, known KBOs have radii of 50--500 km. The measured surface density 
of KBOs on the sky suggests a total mass of $\sim$ 0.1--0.3 \mearth 
(Luu \& Jewitt 1998; Jewitt \etal 1998; Trujillo \etal 2000; 
Kenyon \& Windhorst 2000).  With semimajor axes of 40--50 AU and 
orbital inclinations of 0\degree--30\degree, the apparent geometry 
and mass of the KBO population is similar to results derived for 
the dusty disks observed in $\alpha$ Lyr and $\beta$ Pic, suggesting 
that phenomena currently observed in extra-solar dusty disks probably 
occurred early in the formation of our own solar system.

These discoveries have fueled interest in theoretical studies of
planet formation in a variety of astrophysical contexts.  This paper
reviews some of these studies, with an emphasis on the dynamical 
evolution of planetary debris disks (for other reviews, see chapters 
in Mannings, Boss, \& Russell 2000).  After starting with a short 
summary of current models for planet formation, I consider in 
detail several recent numerical calculations of planet formation in debris 
disks and describe observational tests of these models.  I conclude 
with a discussion of future prospects for the calculations and with 
suggestions for observations to test different models of planet 
formation.

\section{Basic Theory}

\subsection{Background}

Standard models for planet formation begin with a circumstellar
disk surrounding a newly-formed star (Beckwith 1999; Lada 1999).
The disk consists of numerous small dust grains embedded in a 
more massive, gaseous nebula which rotates about the central star.
The disk mass is $\simless$ 10\%--20\% of the mass of the central 
star.  The internal energy of the gas in the disk is also small 
compared to the energy of the gravitational potential;  
the gas thus has a small local scale height compared to 
the distance from the central star. The small velocity shear 
between adjacent disk annuli heats the disk; energy lost to viscous 
heating causes mass to move inwards and angular momentum outwards.  

Two basic theories -- coagulation and dynamical instability -- predict 
that planets can form and survive in this environment.  In the 
coagulation theory, large dust grains within the disk decouple from
the gas and settle 
to the midplane.  These grains may then coagulate into successively 
larger grains (e.g., Weidenschilling 1980; Weidenschilling \& Cuzzi 1993) 
or continue to settle into a very thin layer which can become 
gravitationally unstable (e.g., Goldreich \& Ward 1973).  Both 
paths produce km-sized planetesimals which 
collide and merge to produce larger bodies.  If the growth time is
short compared to the viscous timescale, collisions and mergers
eventually produce one or more `cores' which accumulate most, if not 
all, of the solid mass in an annular `feeding zone' defined by balancing 
the gravity of the growing planetesimal with the gravity of the Sun and 
the rest of the disk.  Large cores with masses of 1--10 \mearth~can 
also accrete gas from the feeding zone (Pollack 1984).  Applied to 
our solar system, this model can account for the masses of the 
terrestrial and several gas giant planets (e.g., Lissauer \etal 1996;
Pollack \etal 1996; Weidenschilling \etal 1997; Levison \etal 1998;
but see Boss 1997, 2000).  Variations of this model, including orbital 
migration and other dynamical processes, are invoked to explain 
Jupiter-sized planets orbiting other solar-type stars (e.g., 
Weidenschilling \& Marzari 1996, Lin \& Ida 1997; Ford \etal 1999; 
Kley 2000).

Dynamical instability models focus on the possibility that part of 
an evolving disk can  collapse directly into a Jupiter-mass planet 
(Boss 1997, 2000; see also Cameron 1995 and references therein).  
If the local surface density is large enough to overcome local 
shear and pressure forces, part of the disk begins to collapse.  
Cool material flows into the growing perturbation, 
which aids the collapse.  Eventually, the perturbation
achieves planet-sized proportions by accumulating all of the gaseous
and solid material in the feeding zone.  This model has a natural
advantage over coagulation models: the collapse is rapid, $\sim$
$10^3$ to $10^5$ yr, compared to the 1--10 Myr lifetime of gaseous 
disks surrounding nearby pre-main sequence stars (Russell \etal 1996;
Hartmann \etal
1998; Lada 1999).  Coagulation models barely succeed in making 
gas giant planets in 1--10 Myr.  However, the disk mass needed to 
start a dynamical instability may exceed the mass typically observed 
in pre-main sequence disks (Beckwith 1999; Lada 1999). Dynamical
instability models produce neither terrestrial planets in the 
inner disk nor icy bodies like Pluto in the outer disk.

\subsection{Disk Physics}

Calculations for both planet formation models require a good
understanding of the important physical processes in the disk.
Gas and solid particles in the disk interact with each other 
and with the gravitational potential and the radiation field of 
the central star.  For the purposes of this review, I ignore 
interactions with external objects, such as nearby stars, molecular 
clouds, and the tidal field of the galaxy.  During the late
stages of planet formation, these interactions are important 
for shaping the evolution of the outermost portions of a solar 
system, at distances of $\sim 10^3$ to $10^4$ AU from the central 
star (see Duncan, Quinn, \& Tremaine 1987, Teplitz \etal 1999, and 
references therein).  External forces can also deliver stochastic
dynamical perturbations to the inner disk, which may aid or inhibit
planet formation depending on the magnitude of the perturbation
(see, for example, Ida \etal 2000; Kalas \etal 2000).

The basic driving mechanism of an isolated gaseous disk is the 
viscosity, which acts to transport mass radially inwards and 
angular momentum radially outwards (von Weisz\"acker 1943, 1948;
see Lin \& Papaloizou 1995, 1996 and Hartmann 1998 for recent reviews).
To understand how this process works,
consider a thin ring with two adjacent annuli at distances 
$r_1$ and $r_2$ from a central star (Figure 1). Material in 
these annuli orbits the central star with velocities, $v_1$ 
and $v_2$. The velocity difference between the two annuli,
$v_1 - v_2 > 0$, produces a frictional force that attempts to equalize 
the two orbital velocities.  The energy lost to friction heats 
the annuli; some disk material then moves inwards to conserve
total energy.  This inward mass motion increases the angular momentum
of the ring; some disk material moves outwards to conserve
angular momentum.  Energy and angular momentum conservation thus 
lead to an expansion of the ring in response to frictional heating.
The ring eventually expands into a disk, which generates heat 
(and radiation) at a level set by the rate mass moves through
the disk, $\mdot$, known as the accretion rate.

In addition to viscous heating, several other processes change the
structure of an accretion disk.  Disks can lose mass and angular 
momentum in a wind, which can be driven by direct radiative 
acceleration or from magnetic energy (e.g., Shu \etal 1994a,b;
Najita \& Shu 1994; Ostriker \& Shu 1995; Proga \etal 1998; 
Proga 2000).  High energy photons and mass loss from the central 
star can heat the disk, raising its energy output 
(Kenyon \& Hartmann 1987; Chiang \& Goldreich 1997, 1999), and
perhaps increasing the rate of mass loss from the wind or from 
evaporation (Hartmann \& Raymond 1984; Johnstone \etal 1998;
Richling \& Yorke 1997, 1998, 2000).  
Gaseous disks are also prone to several types of thermal instabilities, 
which can produce factor of 10--100 changes in the energy output of 
the disk on timescales ranging from minutes to centuries 
(Hartmann \& Kenyon 1985, 1996; Bell \& Lin 1994; Bell \etal 1995; 
Lin \& Papaloizou 1995, 1996; Kenyon 1999).  

\vskip -1ex
\epsfxsize=4.0in
\epsffile{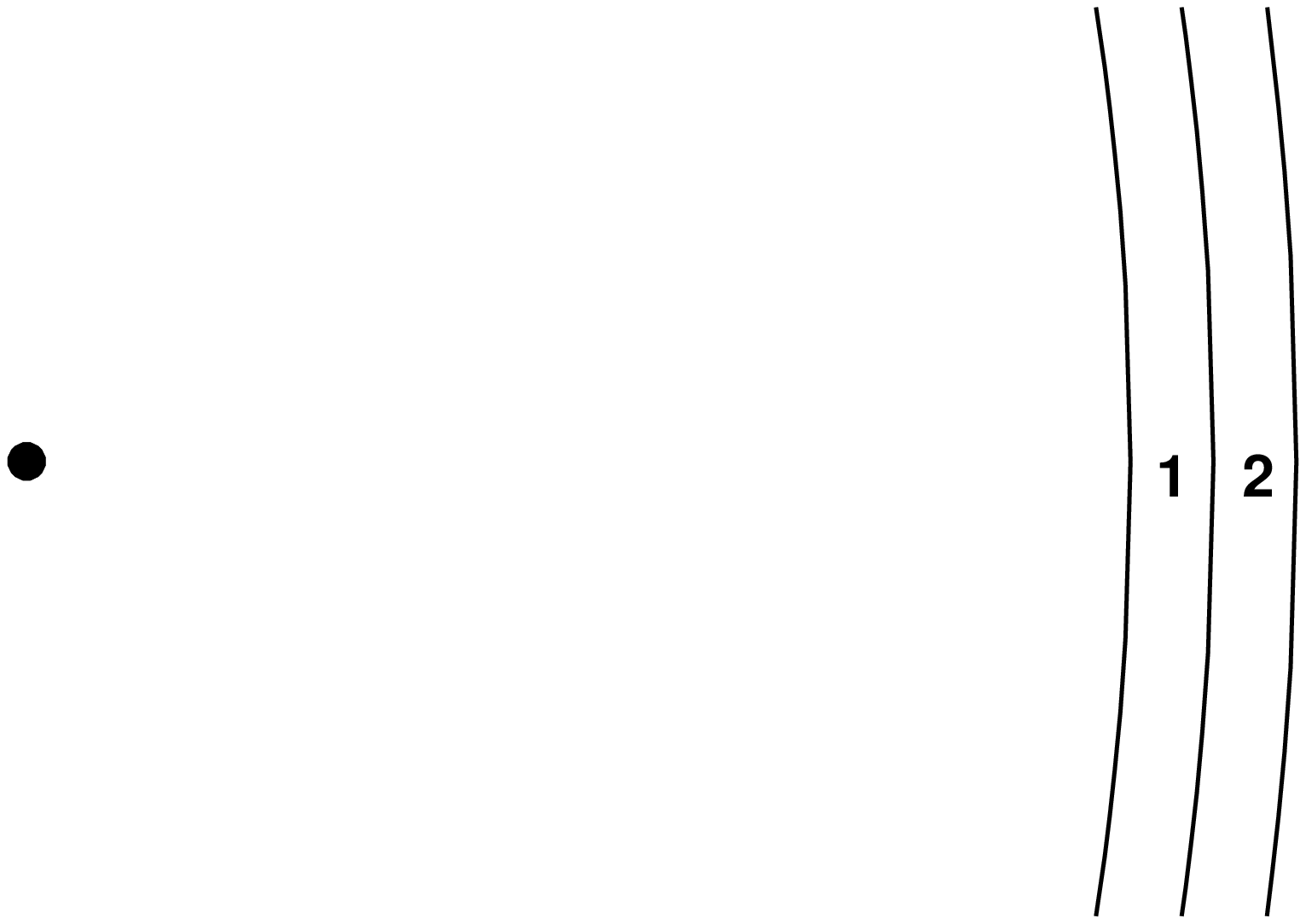}

\footnotesize
\footspace

\vskip -2ex
\noindent{Figure 1 - Schematic view of two adjacent annuli in a disk 
surrounding a compact star.  Annulus `1' lies inside annulus `2'
and orbits the star (filled circle) at a higher velocity,
$v_1 > v_2$.}

\vskip 2ex
\singlespace
\normalsize

The importance of other physical processes in a gaseous disk relative 
to the viscous heating depends on the source of the viscosity, 
which remains controversial.  Ordinary molecular viscosity is 
too small to generate mass motion on a reasonable time scale, 
$\sim$ days for disks in interacting binary systems  and $\sim$ 
years to decades for disks in pre-main sequence stars.  The large 
shear between adjacent disk annuli suggests that disks might be 
unstable to turbulent motions, which has led to many turbulent 
viscosity mechanisms.  Convective eddies, gravitational instabilities, 
internal shocks, magnetic stresses, sound waves, spiral density waves, 
and tidal forces have all been popular turbulence mechanisms 
in the past three decades (see Kenyon 1999 and references therein).

Recent work has shown that magnetic stresses in a differentially
rotating disk inevitably lead to turbulence (Balbus \etal 1996;
Balbus \& Hawley 1998; Stone \etal 2000).
The growth time and effectiveness of these magnetohydrodynamic
mechanisms make them the current leading candidate for viscosity 
in most applications.
How this turbulence leads to significant mass motion in a real
accretion disk remains an unsolved problem.

Shakura \& Sunyaev (1973) avoided some of the basic physical uncertainties 
of viscosity mechanisms with the popular ``$\alpha$-disk'' model
(see also von Weisz\"acker 1943, 1948; L\"ust 1952; 
Lynden-Bell \& Pringle 1974).  They considered viscosity as
a turbulent process, $\nu = \alpha c_s h$, where $c_s$ is the 
sound speed, $h$ is the local scale height of the gas, and $\alpha$ 
is a dimensionless constant. This concept is similar to the mixing 
length theory of convection, with $\alpha$ serving the role of the
mixing length.   In most applications, $\alpha \simless$ 1--10;
$\alpha$ must exceed $\sim 10^{-4}$ to allow material to move 
inwards on a reasonable time scale.

\subsection{Timescales}

This viscosity definition orders the important time scales for gas in 
the disk (Lynden-Bell \& Pringle 1974).  The shortest disk time scale 
is the dynamical (orbital) time scale, which increases radially outward:

\begin{equation}
\tau_D \approx {\rm 1000~yr}~ \left ( \frac{A}{\rm 100~AU} \right )^{3/2} \left ( \frac{1~\msun}{\mstar} \right )^{-1/2}
\end{equation}

\noindent
The thermal time scale measures the rate that energy diffuses through 
the disk,

\begin{equation}
\tau_T \approx {\rm 5000~yr}~ \left ( \frac{A}{\rm 100~AU} \right )^{11/8} ~ .
\end{equation}

\noindent
The thermal time scale is intermediate between the dynamical time scale 
and the viscous time scale, which measures the rate matter diffuses 
through the disk,

\begin{equation}
\tau_V \approx \frac{\rm 25,000~yr}{\alpha} ~ \left ( \frac{A}{\rm 100~AU} \right )^{5/4}  ~ .
\end{equation}

\noindent
These last two expressions do not include a weak dependence on the mass
of the central star.  The unknown $\alpha$ parameter sets the viscous 
timescale; $\tau_V$ is more than 10,000 yr at 100 AU for any reasonable 
$\alpha$.
Most studies of disk evolution suggest $\alpha \sim 10^{-3}$ to $10^{-2}$,
which yields viscous timescales of 1--10 Myr at 100 AU.  Figure 5 of 
Kenyon (1999) shows how the disk reacts to a perturbation in its structure
on each of these timescales.

The final evolutionary timescale for gas in a disk depends on an external
source, the central star, instead of internal physics.
Hollenbach \etal (1994, 2000; also Shu \etal 1993; Richling \& Yorke
1997, 1998, 2000) showed that high 
energy photons from a luminous central star can ionize the outer skin of 
the gaseous disk and raise the gas temperature to $\sim 10^4$ K. The
thermal velocity of this gas is large enough to overcome the local
gravity for material beyond $\sim$ 10 AU for a 1 \msun~central star.
Material then leaves the disk, producing a bipolar outflow which may
be observed in nearby star-forming regions (e.g., Johnstone \etal 1998;
Bally \etal 1998).  Disk evaporation occurs on a timescale 

\begin{equation}
\tau_E \approx {\rm 10^7 ~ yr} \left ( \frac{M_d}{0.01 ~ \msun} \right ) \left ( \frac{A}{\rm 10 ~ AU} \right ) \left ( \frac{\phi_{\star}}{\rm 10^{41} ~ s^{-1}} \right )^{-1/2} ~ ,
\end{equation}

\noindent
where $\phi_{\star}$ is the flux of hydrogen-ionizing photons from
the central star.

The physics of dust evolution in the disk is equally complicated.
Low velocity collisions between dust particles probably produce
larger dust particles; high velocity collisions produce debris.
Radiation pressure can eject small dust grains; gas drag and 
Poynting-Robertson drag can pull small grains into the central 
star if stellar radiation does not evaporate them first.  Although
small dust grains initially orbit the central star in roughly 
circular orbits, drag forces and long-range gravitational perturbations
can modify their orbits significantly over time.  Angular momentum
transfer during elastic collisions -- viscous stirring -- increases
the eccentricity and inclination of all solid bodies in the disk.
Kinetic energy transfer during elastic collisions -- dynamical friction --
leads to ``energy equipartition,'' where the most massive objects 
have more nearly circular orbits than less massive objects
(see Ida \& Makino 1992, 1993).

Despite these complications, we can order the timescales for dust
evolution in the disk.  Ejection of dust grains by radiation pressure
occurs on the local dynamical timescale, equation (1), the shortest
timescale in the disk.  The coagulation timescale is very sensitive
to the particle size and the local particle density, and can range 
from days to months for small grains in a dense medium (see 
Cuzzi \etal 1993; Wurm \& Blum 1998, 2000) 
to many Gyr for large planets in a very low density medium.  
The timescale for grains to settle to the midplane, and the possible
onset of a gravitational instability within this material, depends on
the magnitude and source of turbulence in the disk, of which little
is known.  In current theories, the timescale to produce cm-sized to
km-sized objects is long compared to the local dynamical time but
short compared to the thermal timescale of the gaseous disk (e.g.,
Goldreich \& Ward 1973; Weidenschilling 1980; Weidenschilling \& Cuzzi 
1993; Sekiya \& Ishitsu 2000).  Larger objects are inevitable as long
as the collision velocity in the disk remains low.  The timescale to
make 1000 km objects in a particle disk is:

\begin{equation} 
\tau_{G} \approx {\rm 10^8 ~ yr} \left ( \frac{A}{\rm 100 ~ AU} \right )^{2.5} \left ( \frac{M_d}{\rm 0.01 ~ \msun} \right )^{-1} ~ , 
\end{equation}

\noindent
The growth time is more sensitive to the location in the disk than the 
mass of the disk.  Once 1000 km objects form, larger objects can often 
grow quickly; thus $\tau_G$ is a useful reference for most large planets. 

Two processes prevent coagulation from producing extremely massive objects, 
drag forces and long-range gravitational perturbations.  Gas drag and 
Poynting-Robertson drag remove small objects from the nebula on 
timescales,

\begin{equation}
\tau_{GD} \approx 10^6 ~ {\rm yr} ~ \left ( \frac{m_i}{\rm 10^{9}~g} \right )^{1/3}
         \left ( \frac{A}{\rm 100~AU} \right )^{1/2} 
         \left ( \frac{\rm 1~\msun}{M_{\star}} \right )^{1/2} 
         \left ( \frac{10^{-14}~\rm g~cm^{-3}}{\rho_g} \right )
\end{equation}

\noindent
for gas drag (Adachi \etal 1976; Kary \etal 1993; 10 m objects have
$m_i \sim$ $10^9$ g) and

\begin{equation}
\tau_{PR} \approx {\rm 10^7 ~ yr} ~ \left ( \frac{R}{1 ~ \mum} \right ) \left ( \frac{A}{\rm 100 ~ AU} \right )^2 \left ( \frac{\L_{\star}}{\rm 1 ~ \lsun } \right )^{-1} ~ , 
\end{equation}

\noindent
for Poynting-Robertson drag (Burns \etal 1979).
Gas drag is effective on objects smaller
than $\sim$ 1 km when the gas density in the nebula is large.  
Poynting-Robertson drag is important when high velocity collisions 
produce large amounts of debris in nebulae with little gas content.
The gravitational perturbations viscous stirring and dynamical friction
are important throughout the evolution of a dusty nebula. The timescale
for viscous stirring is comparable to the growth time; dynamical friction
usually acts on a shorter timescale.

Once the largest objects reach their `maximum' size, they rapidly stir up 
the velocities of the smallest objects.   When the velocities are large
enough, collisions tend to produce debris instead of mergers.  Once collisions
produce large amounts of debris, the debris will continue to shatter.  
The timescale to remove bodies from the nebula in this ``collisional 
cascade'' is (Kenyon \& Bromley 2001; see also Backman \etal 1995;
Artymowicz 1997; Lagrange \etal 2000)

\begin{equation} 
\tau_{C} \approx {\rm 5 \times 10^8 ~ yr} \left ( \frac{A}{\rm 100 ~ AU} \right )^{3} \left ( \frac{M_d}{\rm 0.01 ~ \msun} \right )^{-1} ~ .
\end{equation}

With these timescales in place, we can develop a simple understanding
of planet formation in a gaseous disk with some solid material at
the disk midplane (see Figure 2).  This picture depends on observations
of typical sizes, $\sim$ 100--1000 AU, and masses, $\sim$ $10^{-3}$ to
$10^{-1}$ \msun, for circumstellar disks surrounding nearby stars
(see Lagrange \etal 2000; Beckwith, this volume; Koerner, this volume).
For reasonable choices of the viscosity 
coefficient, $\alpha \sim 10^{-3}$ to $10^{-2}$, and the 
ionization rate of the central star, $\phi_{\star} \sim 10^{41}$
s$^{-1}$, the timescale for the disappearance of gas in the disk 
is 1--10 Myr (see equations (3) and (4)).  This timescale is 
similar to the observed upper limit to the lifetime of gaseous disks 
in the solar neighborhood and nearby star-forming regions, 
$\simless$ 10--100 Myr.  

As the gas evolves, dust grains coagulate
and grow to mm or cm sizes on short timescales in the dusty midplane.  
If the dust layer becomes gravitationally unstable, these objects 
can grow to km-sized objects relatively quickly.  Otherwise, 
continuing coagulation gradually produces objects with radii 
of 1 m to 1 km. In either case, coagulation continues to produce 
larger and larger objects.  These objects begin to accrete gas on 
timescales which are very sensitive to the distance from the central 
star, equation (5).
For $M_d \simgreat$ 0.01 \msun~and $A \simless$ 10--20 AU, coagulation 
models can produce gas giant planets before the gas disappears.
Coagulation models currently fail to produce gas giant planets at
larger distances from the central star for any reasonable disk mass
on timescales of 10 Myr or smaller (see Nakagawa \etal 1983;
Fern\'andez \& Ip 1984; Lissauer 1987, 1993, 1995; Ip 1989; 
Pollack \etal 1996).

Once the gas has dispersed, coagulation can continue to produce large
solid objects at $\sim$ 1 AU and at $\simgreat$ 20--30 AU from the 
central star.  Gravitational perturbations from the gas giants
and the largest solid bodies stir up the velocities of the smaller
objects left over from the accumulation process. Collisions between
these small bodies produces debris which is ejected by radiation
pressure, dragged inwards by Poynting-Robertson drag, or incorporated
into larger bodies.  This process continues for the age of the
solar system, although the debris of extra-solar disks probably 
 can be observed for only $\sim$ 1 Gyr (equation 8; 

\vskip -7ex
\epsfxsize=6.0in
\epsffile{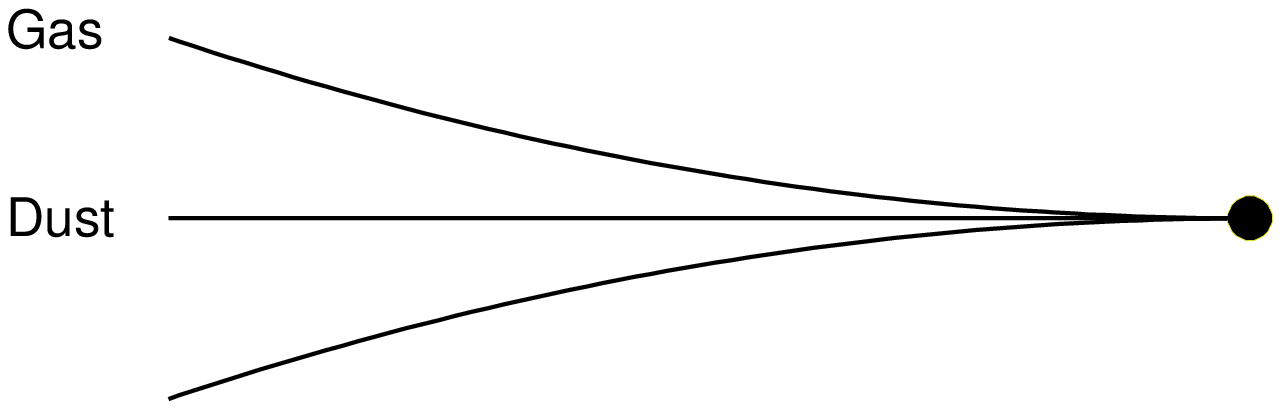}

\footnotesize
\footspace

\vskip -49ex
\noindent{Figure 2 - Schematic side view of a protostellar disk.
Large dust grains settle to the midplane and have a dynamical
scale height, $H_d$, defined by the velocity dispersion about
a circular orbit.  In most cases, $H_d/A \sim$ $10^{-2}$ to $10^{-3}$.
The gas is in vertical hydrostatic equilibrium and has a scale 
height $H_g$ above the midplane; $H_g$ is set by the local gas 
temperature derived from viscous heating and the absorption of 
stellar photons. Gaseous disks usually have $H_d/A \sim$ 0.1--1.0
at $A \sim$ 100 AU (Kenyon \& Hartmann 1987).}

\vskip 6ex
\singlespace
\normalsize

\noindent
debris of extra-solar disks probably 
can be observed for only $\sim$ 1 Gyr (equation 8; Habing \etal 1999;
Lagrange \etal 2000).

To add quantitative details to this picture, we must calculate the
evolution of the gas and dust in a circumstellar disk using a numerical 
model for each physical process.  Various groups have developed 
sophisticated numerical codes to follow the evolution of (i) the gas
(e.g., Ruden \& Pollack 1991; Pollack \etal 1996; Bodenheimer \etal 2000; 
Boss 2000; Kley 2000), (ii) the dust (e.g., Greenberg \etal 1978, 1984;
Wetherill \& Stewart 1989, 1993; Spaute \etal 1991; Weidenschilling \etal 
1997; Kenyon \& Luu 1998, 1999; Kortenkamp \& Wetherill 2000), 
and (iii) the gas and dust (e.g., Bryden \etal 2000; Klahr \& Lin 2001).  
To illustrate the results of these calculations, I will now describe 
coagulation calculations for the outer part of our solar system 
(Kenyon \& Luu 1998, 1999) and the velocity evolution 
of small particles in a debris disk (Kenyon \& Bromley 2001).

\section{Coagulation Calculations}

To calculate the growth of planets from small dust grains,
Safronov (1969) developed the particle-in-a-box method, which
treats planetesimals as a statistical ensemble of masses with 
a distribution of horizontal, $h_i$, and vertical, $v_i$, 
velocities about a circular orbit.  Wetherill \& Stewart (1989,
1993), Spaute \etal (1991), Weidenschilling \etal (1997), and 
Kenyon \& Luu (1998) include references to early work on this 
subject.  A statistical approach is essential, because it is not 
possible to follow the evolution of $10^{15}$ or more small
planetesimals with modern $n$-body codes. Our calculations 
begin with a differential mass distribution, $n(m_i$), in 
concentric annuli centered at heliocentric distances, $A_j$, from 
a star of mass $M_{\star}$ (Kenyon \& Luu 1998, 1999;
Kenyon \& Bromley 2001).  We divide this distribution among $N$ 
mass batches, where $\delta_i \equiv m_{i+1} / m_i$ is the mass
spacing between each batch.  Most calculations have $\delta$ = 1.1--2.0.  
To evolve the mass and velocity distributions in time, we solve the 
coagulation and energy conservation equations for an ensemble of 
objects with masses ranging from $\sim 10^{7}$~g to $\sim 10^{26}$ g.
We adopt analytic cross-sections to derive collision rates, use the
center-of-mass collision energy to infer the collision outcome
(merger, merger + debris, rebound, or disruption), and compute 
velocity changes from gas drag and collective interactions such 
as dynamical friction and viscous stirring using a Fokker-Planck
integrator (Stewart \& Ida 2000; see also Hornung \etal 1985;
Barge \& Pellat 1990; Luciani \etal 1995).  Our code has passed numerous
tests, such as matching the results of analytic solutions to the
coagulation equation (Wetherill 1990; see also Lee 2000; 
Malyshkin \& Goodman 2000) and complete calculations at 1 AU 
(e.g., Wetherill \& Stewart 1993).

\subsection{The Minimum Mass Solar Nebula}

Before describing several coagulation calculations, it is useful
to introduce the `Minimum Mass Solar Nebula,' the minimum amount 
of material needed to build the planets of our solar system (see 
Weidenschilling 1977, Hayashi 1981, Bailey 1994, and references therein).
Hoyle (1946) first emphasized the importance of this concept as a
starting point for theories of solar system formation.  The Minimum 
Mass is based on the close parallel between the measured elemental 
compositions of the earth, Moon, and meteorites and the relative
abundances of heavy elements in the Sun (see the discussion in
Harris 1978). This result leads to the 
assumption that the initial abundance of the solar nebula is
similar to the solar abundance.  The Minimum Mass Solar Nebula
follows from adding hydrogen and helium to each planet to reach a 
solar abundance and spreading the resulting mass uniformly over an
annulus centered on the orbit of the planet.

Figure 3 shows how the mass surface density varies with distance 
for the Minimum Mass Solar Nebula.  The arrows indicate the mass
added to the terrestrial planets.  The plot shows Venus, the Earth, 
Jupiter, Saturn, Uranus, Neptune, and the Kuiper belt.  When the 
material at the orbits of Venus and the Earth is augmented to reach a 
solar abundance of hydrogen, the surface density for the gas follows 
the solid curve, 
$\Sigma_g \approx ~ \Sigma_0 ~ (A / {\rm 1 ~ AU}) ^{-3/2}$, out to 
$A \approx$ 10 AU and then decreases sharply.  The solid curve
in Figure 1 has $\Sigma_0$ = 1500 g cm$^{-2}$; for comparison,
Hayashi \etal (1985) proposed $\Sigma_0$ = 1700 g cm$^{-2}$ while
Weidenschilling (1977) suggested $\Sigma_0$ = 3200 g cm$^{-2}$.
Following Hayashi (1981), the dot-dashed curve has

\begin{equation}
\Sigma_s = \left\{ \begin{array}{l l l}
	   7 ~ {\rm g ~ cm^{-2}} ~ (A / {\rm 1 ~ AU}) ^{-3/2} & \hspace{5mm} & A \le 2.7 ~ {\rm AU} \\
                     \\
	   30 ~ {\rm g ~ cm^{-2}} ~ (A / {\rm 1 ~ AU}) ^{-3/2} & \hspace{5mm} & A > 2.7 ~ {\rm AU} \\
 \end{array}
         \right .
\end{equation}

\vskip 2ex
\noindent
The uncertainties in the coefficients are again a factor of $\sim$ 2.  The 
change in the surface density of solid material at 2.7 AU corresponds to 
the region where ice condenses out of the gas in Hayashi's (1981) model.
The location of this region depends on the disk structure (Sasselov \&
Lecar 2000).

The Minimum Mass Solar Nebula was one of the great successes of
early viscous accretion disk theories, because steady-state disk
models often yield $\Sigma \propto A^{-3/2}$.  The sharp decrease
in $\Sigma$ at 10--30 AU supports photoevaporation models, 
because ionized hydrogen becomes unbound at $\sim$ 10 AU 
(Shu \etal 1993).  Current abundance measurements for the gas 
giants lend additional evidence: the He/H ratio appears to decrease

\vskip -8ex
\epsfxsize=5.5in
\epsffile{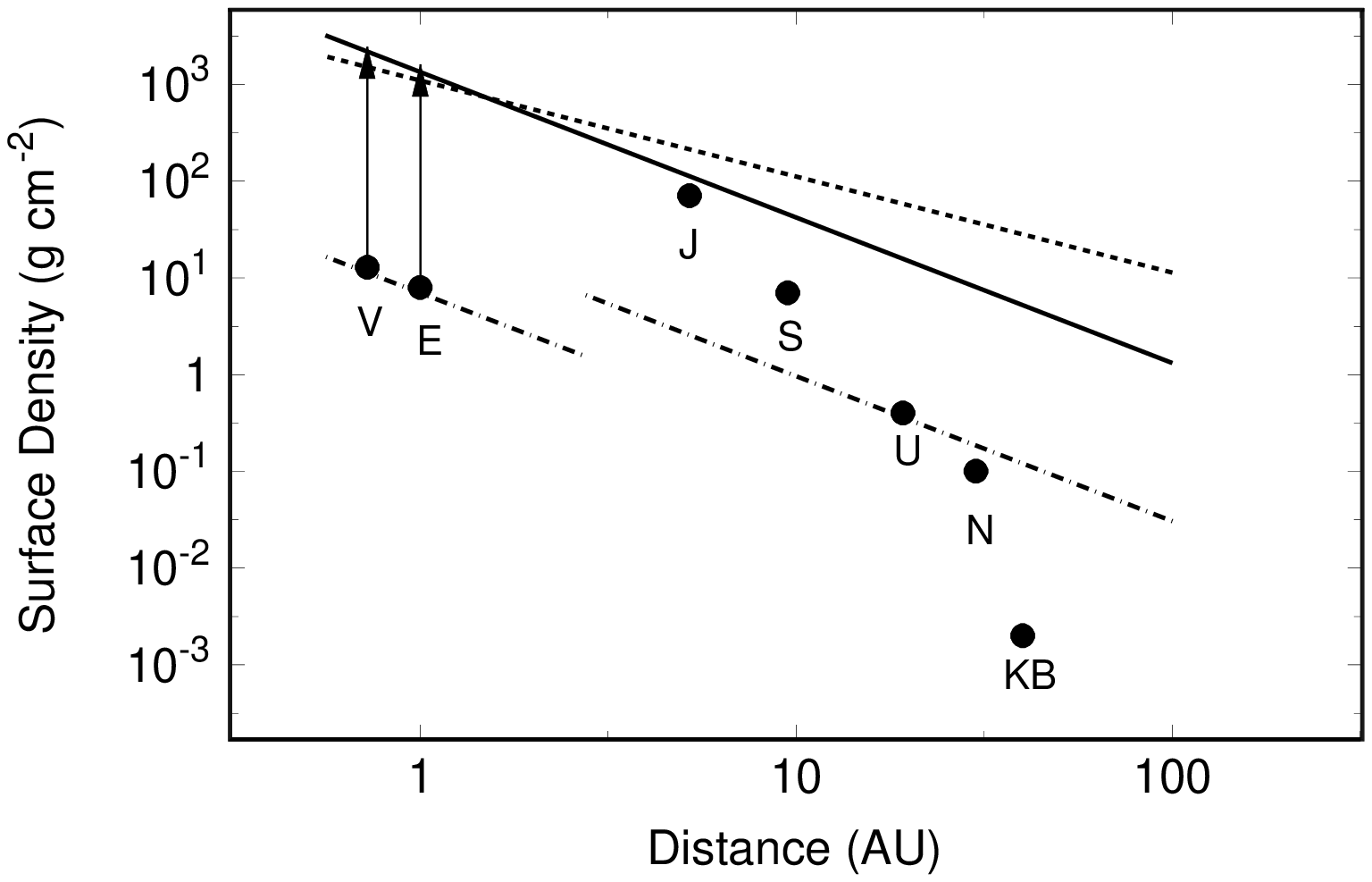}

\footnotesize
\footspace

\vskip -10ex
\noindent{Figure 3 - Surface density distribution in the solar system,
assuming that the mass is spread uniformly over an annulus centered on 
the orbit of the planet.  The arrows indicate the surface density
for terrestrial planets if augmented to a solar hydrogen abundance.
The solid and dot-dashed curves indicate $\Sigma \propto A^{-3/2}$;
the dashed line indicates $\Sigma \propto A^{-1}$ (Cameron 1995).}

\vskip 5ex
\singlespace
\normalsize

\noindent
1993).  Current abundance measurements for the gas giants lend 
additional evidence: the gas-to-dust ratio appears to decrease
with radius in parallel with the surface density drop beyond 10 AU
(Pollack 1984; Podolak \etal 1985; Podolak \& Reynolds 1987;
Pollack \etal 1996).  In the Kuiper Belt, 
there may be two origins for the 
additional large drop in surface density from a 
$\Sigma \propto A^{-3/2}$ model.  Adding H and He at 30--40 AU
increases the mass in the Kuiper Belt by a factor of $\sim$ 30. 
Material
lost to high velocity collision of objects in the Belt increases
the mass by another factor of 10--100 (e.g., Holman \& Wisdom 1993;
Davis \& Farinella 1997; Kenyon \& Luu 1999a, b), bringing the
initial surface density in the Kuiper Belt within range of the
$\Sigma \propto A^{-3/2}$ line.  If these estimates are correct,
the total mass of the Minimum Mass Solar Nebula is $\sim$ 0.01 \msun~for
an outer radius of $\sim$ 100 AU, close to the median mass for 
circumstellar disks surrounding young stars in nearby regions 
of star formation (Lada 1999).

Figure 4 suggests that the Kuiper Belt provides an important test 
of coagulation models.  Forming objects with radii of $\sim$ 
500--1000 km requires $\sim 10^7$ yr at $\sim$ 40 AU in a Minimum 
Mass Solar Nebula (equation (5)).  The outermost gas giant, Neptune,
must form on a similar timescale to accrete gas from the solar nebula 
before the gas escapes (equation (3)).  Neptune formation places 
another constraint on the KBO growth time, because Neptune inhibits 
KBO formation at 30--40 AU by increasing particle random 
velocities on timescales of 20--100 Myr (Holman \& Wisdom 1993;
Duncan \etal 1995; Morbidelli \& Valsecchi 1997).  Kenyon \& Luu
(1998, 1999a,b) investigated how KBOs form by coagulation and
compared their results with observations (see also Fern\'andez 1997;
Stern \& Colwell 1997a, b and references therein).  The next section briefly 
describes the model results; \S4 compares these results with observations.

\subsection{Kuiper Belt Models}

Figure 4 shows the results of a complete coagulation calculation for the 
Kuiper Belt in our solar system.  The input cumulative size distribution 
$N_C$ is $N_C \propto r_i^{q_0}$, with initial radii $r_i$ = 1--80 m and 
$q_0$ = 3. We assume these particles are uniformly distributed
in a single annulus with a width of 6 AU at 32--38 AU from the Sun.
The total mass in the annulus is $M_0$; $M_0 \approx$ 10 \mearth~for 
a Minimum Mass Solar Nebula.  All mass batches start with the same 
initial velocity.  We tested a range of initial velocities 
corresponding to initial eccentricities of $e_0 = 10^{-4}$ to 
$10^{-2}$, as is expected for planetesimals in the early solar 
nebula (Malhotra 1995).  The adopted mass density, $\rho_0$ = 
1.5 g cm$^{-3}$, is appropriate for icy bodies with a small 
rocky component.  Kenyon \& Luu (1999a,b) describe these parameters 
in more detail.

\epsfxsize=5.3in
\epsffile{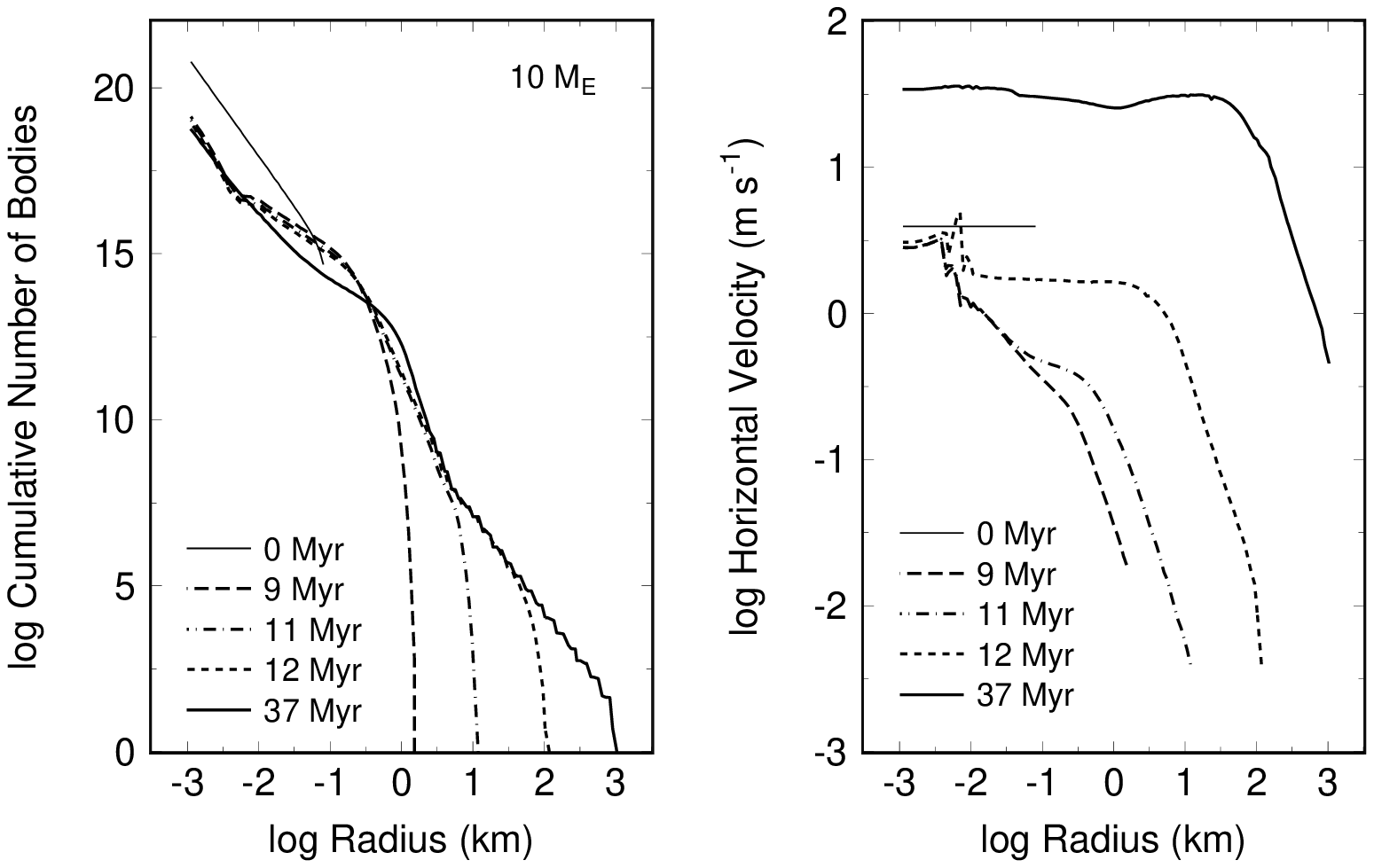}

\footnotesize
\footspace

\vskip -11ex
\noindent{Figure 4 - Cumulative size distributions as a function
of time for a model with $M_0 = 10 \mearth$ from Kenyon \& Luu (1999a).  
The evolution time for each curve is listed in the legend. 
Crosses indicate observational and theoretical constraints 
on the size distribution at radii of 50 km, 10 km, and 1 km
as described in the text.}

\vskip 6ex
\singlespace
\normalsize

We separate the growth of KBOs into three regimes.  Early in the 
evolution, frequent collisions damp the velocity dispersions of small 
bodies.  These bodies slowly grow into 1~km objects on a timescale 
of 5--10 Myr $(M_0/10 ~ \mearth)^{-1}$.  This linear growth phase ends when 
the gravitational range of the largest objects exceeds their 
geometric cross-section.  This ``gravitational focusing'' enhances 
the collision rate by factors of 10--1000.  The largest objects 
then begin a period of ``runaway growth'', when their radii grow from 
$\sim$ 1~km to $\gtrsim$ 100~km in several Myr.  During this phase, 
dynamical friction and viscous stirring increase the velocity dispersions 
of the smallest bodies from $\sim$ 1~m~s$^{-1}$ up to $\sim$ 40~m~s$^{-1}$.
This velocity evolution reduces gravitational focusing factors and ends 
runaway growth.  The largest objects then grow slowly to 1000+ km sizes 
on timescales that again depend on the initial mass in the annulus.
Kokubo \& Ida (1998) call this last phase in the evolution 
`oligarchic growth' to distinguish it from the linear and 
runaway growth phases (see also Ida \& Makino 1993.

The shapes of the curves in Figure 4 show features common to all
coagulation calculations.  Almost all codes produce two power-law 
size distributions, $N_C \propto r_i^{-q_f}$. The merger component 
at large sizes has $q_f \approx$ 3; the debris component at small 
sizes has $q_f = 2.5$ (Dohnanyi 1969).  Dynamical friction produces 
a power law velocity in the merger component.  Because the debris 
component contains a small fraction of the initial mass, it has roughly 
constant velocity.  The transition region between the two components 
usually has a `bump' in the size distribution, where objects which 
can merge grow rapidly to join the merger population.  Calculations 
for annuli closer to the Sun also yield a `runaway' population, a 
plateau in the size distribution of the largest objects.

Our Kuiper Belt calculations yield one result which is very different
from coagulation calculations for annuli at less than 10 AU from the
Sun.  In all other published calculations, the largest bodies contain
most of the initial mass in the annulus.  In the Kuiper Belt, most of
the initial mass ends up in 1 km objects.  Fragmentation and gravitational
stirring are responsible for this difference between calculations at
1--10 AU and at 40 AU.  In our calculations, fragmentation produces a 
large reservoir of small bodies that damp the velocity dispersions of 
the large objects through dynamical friction.  These processes allow 
a short runaway growth phase where 1 km objects grow into 100 km objects.  
Continued fragmentation and velocity evolution damp runaway growth by 
increasing the velocity dispersions of small objects.  Our models thus
enter the phase of `oligarchic growth' earlier than models for planet 
growth at 1--10 AU.  This evolution leaves $\sim$ 1\%--2\% of the 
initial mass in 100--1000 km objects.\footnote{Calculations by Stern (1995,
1996) and Stern \& Colwell (1997a,b) do not include a self-consistent
algorithm for velocity evolution and thus do not achieve these results.}
The remaining mass is in 
0.1--10~km radius objects.  Continued fragmentation will gradually 
erode these smaller objects into dust grains that are removed from 
the Kuiper Belt on short timescales, $\sim 10^7$ yr (see Backman
\& Paresce 1993; Backman \etal 1995). Thus, in our interpretation, 
100--1000 km radius objects comprise a small fraction of the original 
Kuiper Belt.  This conclusion supports the notion that the solar nebula
followed the $\Sigma \propto A^{-3/2}$ law out to $\sim$ 40--50 AU.

The coagulation calculations for the Kuiper Belt also support a link
between the formation of large icy objects like KBOs and the evolution of 
dusty debris disks in $\alpha$ Lyr, $\beta$ Pic, and other main sequence 
stars.  In our models, the growth of 100--1000 km radius KBOs in the
outer solar system is accompanied by substantial dust production, 
$\sim$ 0.1--1 $M_E$, in models with initial surface densities of
1--5 times the Minimum Mass Solar Nebula.  This dust mass is comparable 
to the `maximum' masses inferred for $\alpha$ Lyr and $\beta$ Pic
(e.g., Backman \& Paresce 1993; Lagrange \etal 2000).  The opening angle
of the disk in $\beta$ Pic implies vertical velocity dispersions of
$\sim$ 100 m s$^{-1}$, close to that derived for the Kuiper Belt
calculations.  To explore this connection in more detail, we have
begun a set of coagulation calculations appropriate for debris disks
surrounding nearby A-type stars. The following section describes our
progress.

\subsection{Debris Disk Models}

Current models for a debris disk envision a collection of dust
grains in circular orbits around an A-type star (see Artymowicz
1997; Lagrange \etal 2000).  For particle sizes of 1--100 \mum, 
Poynting-Robertson drag and radiation pressure remove dust from 
the disk in $\sim $ 1--10 Myr (equation (7)).  Collisions between 
larger bodies can replenish small grains if the collision velocity 
is $\sim$ 100--300 m s$^{-1}$.  These large velocities initiate a 
``collisional cascade,'' where planetesimals with radii of 1--10 km 
are ground down into smaller and smaller bodies (Artymowicz \etal 1989;
Backman \etal 1995).  A collisional 
cascade requires a mass reservoir of $\sim$ 10--100 $\mearth$ to 
replenish smaller grains over a disk lifetime of 100 Myr or more
(Artymowicz 1997; Habing \etal 1999; Lagrange \etal 2000;
Kenyon \& Bromley 2001).

The origin of the large collision velocities in this picture is 
uncertain.  Because the difference between the gas velocity and 
dust velocity is small and circularization is efficient, dust grains 
and larger bodies within a protosolar nebula probably had nearly 
circular orbits initially.  Short-term encounters with passing stars 
and stirring by planets embedded in the disk can increase the 
velocities of dust grains (Artymowicz \etal 1989; Larwood 1997;
Mouillet \etal 1997;
Ida \etal 2000; Kalas \etal 2000).  Although stellar encounters 
can increase particle velocities enormously, such encounters are 
probably rare.  Collisions of bodies within the disk may also 
effectively damp large velocities following the encounter; 
Kenyon \& Luu (1999a) estimate damping times of 1--10 Myr 
for 1--100 m objects during the early stages of planetesimal 
growth in the Kuiper Belt.  Stirring by embedded planets is 
attractive, because objects with radii of 1000 km or more 
naturally form in the outer disk (\S3.2) and these objects 
continuously stir up the velocities of small dust grains
(see also Kenyon \etal 1999).

Kenyon \& Bromley (2001) recently examined the possibility that 
large objects embedded in a debris disk can stir up the velocities 
of smaller objects to the `shattering limit' and thus produce a 
collisional cascade (see also Backman \& Paresce 1993; Backman 
\etal 1995; Artymowicz 1997; Kenyon \etal 1999; Lagrange \etal 2000).  
Using reasonable physical constants for the grain population, 
they show that 1 km objects with collision velocities 
of 100 m s$^{-1}$ or more can replenish the small grain population
in a debris disk.  Because gravitational stirring by small objects
is inefficient, a disk composed of 1 km objects, can {\it never} 
reach the 100 m s$^{-1}$ limit in 1 Gyr or less.  
Disks composed of 100 km or larger objects can reach the 
100 m s$^{-1}$ limit in less than 100 Myr, but these objects 
cannot be shattered.  Kenyon \& Bromley (2001) proposed that
the shattering limit might be reached more easily in a disk 
containing a size distribution of planetesimals.  They assumed
an initial size distribution similar to the results derived from
the complete coagulation models of Kenyon \& Luu (1999a, b),
$N_C \propto r_i^{-q_f}$ with $q_f$ = 3.  They then varied the 
maximum mass of this distribution to derive limits on the stirring
as a function of mass.

\vskip 2ex
\epsfxsize=6.6in
\epsffile{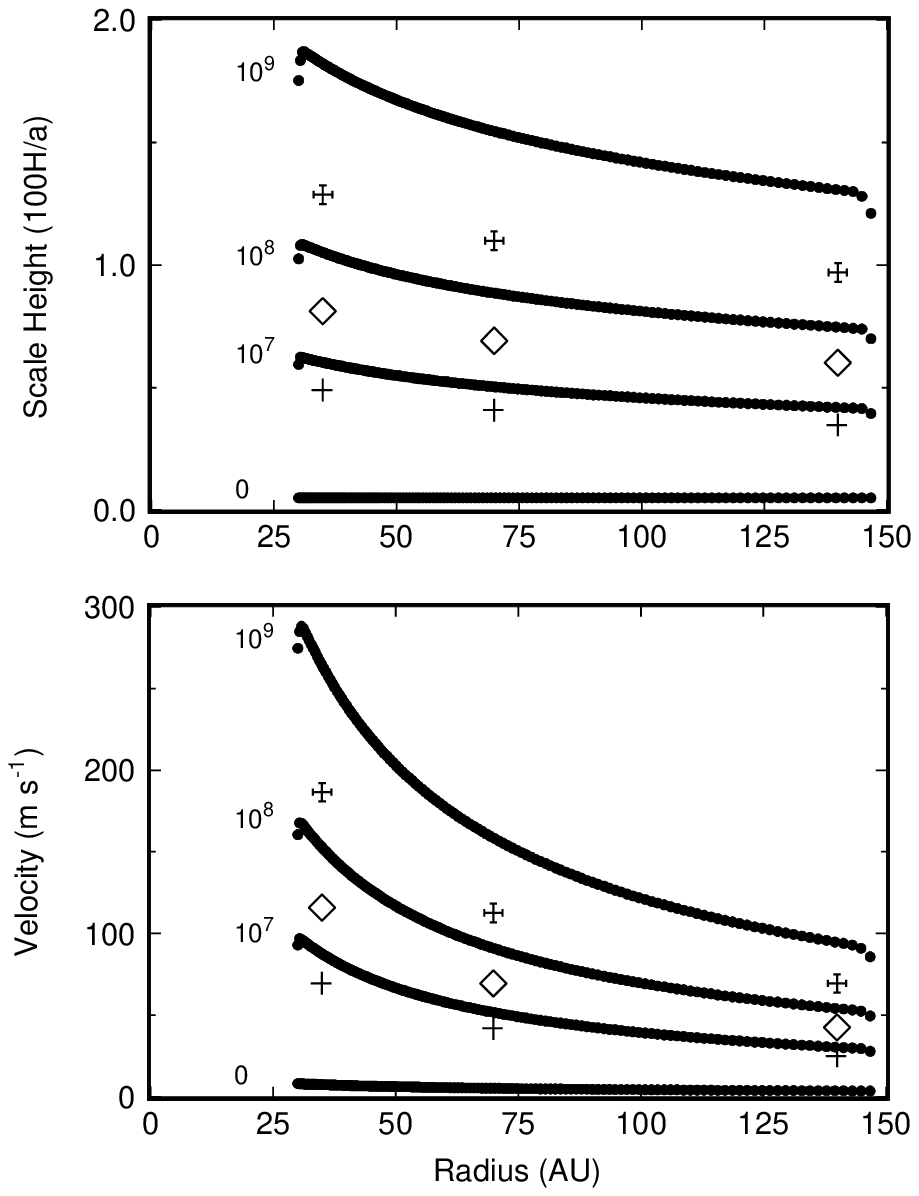}

\footnotesize
\footspace

\noindent{Figure 5 - Evolution of particle velocity (left panel) and 
vertical scale height (right panel) for 1 m to 1 km planetesimals of a size 
distribution of 1 m to 500 km planetesimals in a disk surrounding a 
3 \msun~star.  The initial surface density in the disk is $\Sigma = 
60 ~ (a/{\rm 1 ~ AU})^{-3/2}$.  The evolution time in years is listed 
to the left of each curve. The symbols indicate results for partial
disk calculations composed of 16 annuli, instead of the 128 annuli
used in the full disk calculation.}

\vskip 6ex
\singlespace
\normalsize

Figure 5 shows the evolution of particle velocity (lower panel) and 
vertical scale height (top panel) for small planetesimals of a size 
distribution of 1 m to 500 km planetesimals in a disk surrounding a 
3 \msun~star.  The initial surface density in the disk is $\Sigma = 
60 ~ (a/{\rm 1 ~ AU})^{-3/2}$, somewhat larger than the surface density 
of a Minimum Mass Solar Nebula.  The evolution time in years is listed 
to the left of each curve.  The filled circles show the evolution for 
a complete disk; the symbols indicate the slower evolution for 
isolated annuli.  Distant gravitational perturbations between large 
bodies are an important source of velocity evolution (see also
Weidenschilling 1989; Stewart \& Ida 2000); the evolution of 
the particle velocity in isolated annuli thus lags results for  
a complete disk model.  These results demonstrate that a few 500 km 
bodies can stir the velocities of 1 m to 1 km bodies in a circumstellar 
disk to the shattering limit of 100 m s$^{-1}$ or larger on timescales 
of $10^7$ or more years.  The scale height $H_d$ of the dust reaches 
the observable limit of 100 $H_d/A$ = 1, on a similar timescale. In 
contrast, stirring by 100 km bodies is ineffective on 1 Gyr timescales
(see Figure 6 of Kenyon \& Bromley 2001).  Thus, if large bodies
embedded in a planetesimal disk are responsible for the large observed
velocities of small bodies, these calculations show that the large
bodies must have radii of 500 km or larger. 

Although these calculations are too new for detailed comparisons with
observations, observed scale height profiles of debris disk systems 
will eventually place useful constraints on the models.  The scale 
height distribution of $H \propto r^{3/4}$ in Figure 5 is shallower
than the typical distribution observed in $\beta$ Pic and other systems
$H \propto r$ (e.g., Artymowicz 1997; Lagrange \etal 2000).  Radiation
pressure on shattered grains in the inner portions of the disk should
increase the scale height at larger radii; Poynting-Robertson drag on
shattered grains in the outer parts of the disk should decrease the
scale height at small radii.  Coagulation tends to produce larger
objects in the inner portions of the disk and should produce larger
scale heights in the inner disk.  Future calculations will allow us
to see how these competing physical processes act on the observed 
scale height.  However, we are encouraged that the scale height
distribution derived solely from the stirring calculations is close
to those observed in real debris disk systems.

\subsection{Summary}

The particle-in-a-box formulation works well during the early stages 
of planet formation. When the number of small planetesimals is large,
the distribution of bodies is well-approximated by a uniform density
in a narrow grid; the collision rates and velocity evolution can
then be treated using fairly simple statistical formulae.  More
general $n$-body calculations confirm the basic features of 
particle-in-a-box calculations for the early stages of planet growth 
described here (Ida \& Makino 1992; Kokubo \& Ida 1996).  The method 
begins to break down in the late stages of planetary growth, when
the number of large planets is small and one-on-one interactions
are important.  Direct $n$-body calculations are then more accurate,
but these models cannot follow the numerous small bodies left over
from the growth of a few very large objects (e.g., Chambers \& 
Wetherill 1998; Nagasawa \etal 2000).  Because small objects can contain 
a substantial fraction of the initial disk mass, they make 
important contributions to the dynamical evolution of the
largest objects throughout the evolution.  Several groups are now 
trying to merge coagulation calculations with direct $n$-body codes 
(see, for example, Weidenschilling \etal 1997).  These hybrid codes 
should allow better calculations of planet formation from 
``start to finish.''

\section{Tests of Coagulation Models}

Most coagulation codes have numerous free parameters which are poorly
constrained by theory.  For example, algorithms to compute the debris
produced in a collision between two or more bodies require as input the
tensile strength, $S_0$; the gravitational binding energy, $S_g$; the
velocity of the ejecta as a function of mass, $f(>v)$; and the number
distribution of the ejecta, $n(M)$.  Some of these parameters can be
constrained by laboratory experiments on Earth, but some cannot.  Until
very recently, our solar system has been the only external laboratory 
to test models. Aside from obvious tests of calculations to
attain the masses of the planets in our solar system, comparisons
between calculations and solar system data yield constraints on the
bulk properties of asteroids and other rocky objects in the inner
solar system (e.g., Davis \etal 1985, 1994) and on physical processes 
such as gas drag and Poynting-Robertson drag (see, for example, Burns
\etal 1979, Kary \etal 1993, and references therein).  Here we describe 
several tests of the coagulation code based on objects in the outer
solar system.

Kenyon \& Luu (1999b) showed that complete coagulation calculations 
for the outer solar system yield a power law number distribution,

\begin{equation}
{\rm log} ~ N_C = N_0 \left ( \frac{m}{m_0} \right )^{-a} ~ ,
\end{equation}

\noindent
where $N_C$ is the cumulative number of bodies with mass larger
than $m$.  Both $m_0$ and $N_0$ are scaling parameters which 
depend on the initial mass in the disk.  The power law exponent,
$a$, has a small range for the largest bodies -- $a = 4.00 \pm 0.25$ -- 
which is fairly
independent of the initial conditions and uncertain physical parameters 
for fragmentation and other processes. This result leads to a direct
prediction for the observed luminosity function (LF) of bodies
in the Kuiper Belt beyond the orbit of Neptune:

\begin{equation}
{\rm log} ~ N = \alpha (R - R_0) ~ ,
\end{equation}

\noindent
where $N$ is the cumulative number of objects brighter than magnitude 
$R$ and $R_0$ is a reference magnitude.  Coagulation models with $a = 4$
predict $\alpha$ = 0.6; observations yield $\alpha$ = 0.5--0.7.

Figure 6 compares model luminosity functions of KBOs
with observations.  Data are as indicated in the legend of each panel.
The open circle with the central dot is the position of Pluto for an
adopted albedo of 4\%; other observations are from the literature
(see Kenyon \& Luu 1999b).  Error bars for 
each datum -- typically a factor of 2--3 -- are not shown for clarity. 
The lines plot luminosity functions for models 
with (a) left panel: an initial eccentricity $e_0 = 10^{-3}$ for
planetesimals with radii of 1--100 m and $M_0 \approx$ 0.3 (dot-dashed), 
1.0 (solid), and 3.0 (dashed) times the Minimum Mass Solar Nebula and 
(b) right panel: a Minimum Mass Solar Nebula with $e_0 = 10^{-2}$ 
(dashed), $e_0 = 10^{-3}$ (solid), and $e_0 = 10^{-4}$ (dot-dashed).  
The pair of vertical solid lines indicates the planned magnitude range 
accessible to {\it NGST}. The model luminosity functions agree well
with current observations.

The good agrement between models and observations for R = 20--26 in 
Figure 6 is encouraging. The major uncertainties is this comparison --
the orbit distribution of KBOs and the evolution of the luminosity
function with time -- are difficult to quantify.  As ground-based
telescopes reveal more KBOs, the improved distribution of KBO orbits 
will yield a more reliable comparison with the models.  Collisional
erosion over the age of the solar system probably does not affect 
large KBOs (Davis \& Farinella 1997), but dynamical encounters with
Neptune can remove large KBOs from the population.  This process cannot
change the slope of the luminosity function, but it can change the
scaling parameter by a factor of 2 or more (Holman \& Wisdom 1993).

Other tests of KBO coagulation models can be made with other data.  
In our models, disruption of colliding planetesimals can prevent 
the growth of large objects if the intrinsic tensile strength of
the bodies is small (Kenyon \& Luu 1999a).  If Pluto formed by 
coagulation, the minimum tensile strength for Kuiper Belt objects 
in the outer solar system is $S_0 = 300$ erg g$^{-1}$ (see Ryan \etal
1999 for laboratory experiments appropriate for collisions among
Kuiper Belt objects).  The debris
produced by colliding planetesimals should follow a size distribution
with $q_f \approx$ 3.5, somewhat shallower than the slope of the
merger population, $q_f \approx -4$ (Dohnanyi 1969; Davis \& 
Farinella 1997). Our models predict that this transition should
occur for objects with radii of $\sim$ 1 km (Kenyon \& Luu 1999a).  
Kenyon \& Windhorst (2000) used Olbers Paradox and the measured 
brightness of the optical night sky to show that the faint end of 
the size distribution for Kuiper Belt objects has $q_f \simless -3.5$
for objects with sizes of 1 m to 1 km.  Observations with {\it NGST} 
can search for a break in the slope of the size distribution at
1 km (R $\approx$ 30), as illustrated in Figure 6.

\epsfxsize=5.5in
\epsffile{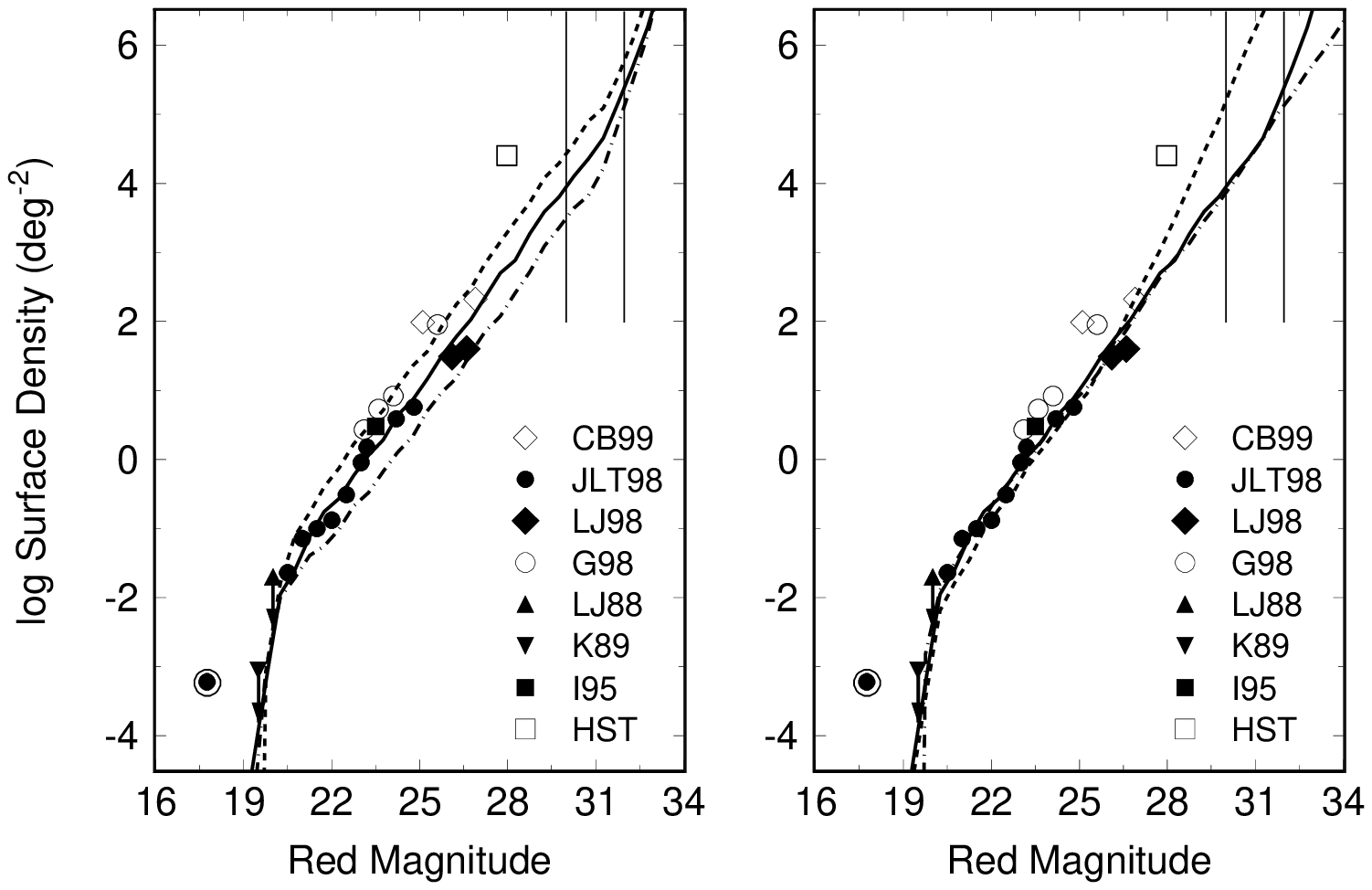}

\footnotesize
\footspace

\vskip -11ex
\noindent{Figure 6 - 
Comparison of model luminosity functions of KBOs
with observations.  Data are as indicated in the legend of each panel.
The open circle with the central dot is the position of Pluto for an
adopted albedo of 4\%; other observations are from 
Cochran \etal (1998; HST), Irwin \etal (1995; I95), 
Kowal 1989 (1989; K89), Luu \& Jewitt (1988; LJ88),
Gladman \etal (1998; G98), Luu \& Jewitt (1998; LJ98), 
Jewitt \etal (1998; JLT98) and Chiang \& Brown (1999; CB99).  
Error bars for each datum -- typically a factor of 2--3 -- and 
the upper limit from Levison \& Duncan (1990) are not shown 
for clarity. The lines plot luminosity functions for models 
with (a) left panel: $e_0 = 10^{-3}$ and $M_0
\approx$ 0.3 (dot-dashed), 1.0 (solid), and 3.0 (dashed) times the
Minimum Mass Solar Nebula and (b) right panel: a Minimum Mass Solar
Nebula with $e_0 = 10^{-2}$ (dashed), $e_0 = 10^{-3}$ (solid), and
$e_0 = 10^{-4}$ (dot-dashed).  A Minimum Mass Solar Nebula has
$M_0 \approx 12~\mearth$ within R = 42--50 AU.  The pair of vertical solid 
lines indicates the planned magnitude range accessible to {\it NGST}.

\vskip 6ex
\singlespace
\normalsize

Observations of debris disk systems provide other measurements of
the slope of the size distribution for small objects.  The data
are consistent with the predicted slopes, but the uncertainties
remain large (see Kenyon \etal 1999; Lagrange \etal 2000).  

In addition to the geometry and mass estimates derived from images 
and spectral energy distributions, many observations of debris disk 
systems have concentrated on measurements of the composition and
other bulk properties of solid objects (see Pantin \etal 1997,
Lagrange \etal 2000, and references therein).  Many systems show 
evidence for infalling `comets' and other gaseous phenomena in
the disk (e.g., Beust \etal 1998, Grady \etal 1999, Sitko \etal 
1999, Lagrange \etal 2000, and references therein).  Although 
coagulation calculations do not address directly questions
concerning the composition of solid bodies in debris disks, 
calculations will eventually yield accurate rates of dust 
infall and ejection for comparison with observations. 

\section{Summary}

Understanding whether or not coagulation or dynamical instability 
produce solar systems similar to those we observe involves a
comparison of detailed numerical calculations with high quality
observations.  The advent of fast, parallel computers now allows
fairly realistic calculations of the complete evolution of part 
of a solar system; calculations of an entire solar system will
be possible in 10 years or less.  Observations with ground-based
and space-based telescopes now provide data with sufficient 
sensitivity and spatial resolution to make direct tests of some
predictions of the detailed numerical calculations.  These tests
yield a better understanding of the physical processes involved
in forming terrestrial planets like the Earth and gas giant planets
like Jupiter.  As both the data and models improve, we will have
a better grasp of the origin of planetary systems similar to our own.

In addition to the tests described in \S4, ground-based and space-based
observations will yield new tests of planet formation theories.
Data from {\it SIRTF}, {\it SOFIA}, and other NASA-sponsored missions 
will result in many new debris disk systems and better estimates for 
the distribution of lifetimes for these systems.  Comparisons with the 
results of the coagulation calculations will test theories of the
collisional cascade.  Direct detection of more large Kuiper Belt
objects will test predictions for the merger population; direct 
detection of the optical and far-infrared background from KBOs will 
test predictions for the size distribution of the debris population 
and a measurement of the albedo of KBOs (Kenyon \& Windhorst 2000).
These comparisons -- combined with new calculations for the Kuiper 
Belt as in Kenyon \& Luu (1998, 1999) -- will give us a better 
understanding of planet formation (for Earth-sized planets) in 
the outer part of a solar system.

Along with simple observational comparisons of the disk lifetime,
inclusion of good models for gas drag, Poynting-Robertson drag,
and radiation pressure on grains will allow predictions for the
surface brightness profile of solar system disks, derived from
the model results in Figure 5.  The {\it HST} and several 
ground-based telescopes have derived observed surface brightness 
profiles; the {\it NGST} mission will provide many hundreds.  
Comparisons of observations with coagulation model predictions 
will allow better constraints on the models and on the physical 
situation in extra-solar debris disks.  The identification of features
in the observed profiles may yield constraints on planetary masses
in the disks.

Observations may also help guide theorists in producing better models
for gas giant planets.  Current theory has trouble understanding the
formation of gas giants at distance of $\simgreat$ 10 AU (coagulation 
models) and $\simgreat$ 20 AU (dynamical instability models) from the
central star. Identification of gas giant planets in other solar systems
at large distances from their parent star would provide a stern test
of either model.  Discovery of gas giants in the weak-emission T Tauri
stars (wTTs) might be the sternest test of all. Many wTTs have little 
or no evidence for emission from disk material at $A \simless$ 1--10
AU, suggesting that planet formation may be complete in these systems.
A gas giant planet discovered beyond 20 AU from the central star in a
wTTS would probably be luminous due to its young age and might yield
important insights into the formation of gas giant planets.

\vskip 6ex

I would like to thank Jane Luu.  Without Jane's foresight and persistence,
our joint projects on planet formation would never have been possible.
I also thank B. Bromley, A. Cameron, F. Franklin, M. Geller, M. Holman, 
and J. Wood for advice and comments. I also acknowledge generous 
allotments of computer time on the JPL Cray T3D `Cosmos', the HP 
Exemplar `Neptune' and the Silicon Graphics Origin-2000 `Alhena' 
through funding from the NASA Offices of Mission to Planet Earth, 
Aeronautics, and Space Science.

\end{document}